\documentclass[letterpaper,twocolumn,pr,aps,superscriptaddress,eqsecnum,nofootinbib]{revtex4}

\usepackage{amssymb}
\usepackage{amsmath}
\usepackage{amsthm}
\usepackage{cancel}
\usepackage{framed}
\usepackage{enumerate}
\usepackage{color}
\usepackage{bbm}
\usepackage{mathrsfs}
\usepackage[bookmarks=false]{hyperref}

\newtheorem{theorem}{Theorem}

\theoremstyle{definition}

\newcommand*{\bbN}{\mathbb{N}}
\newcommand*{\bbR}{\mathbb{R}}

\newcommand*{\bbZ}{\mathbb{Z}}

\newcommand*{\cD}{\mathcal{D}}
\newcommand*{\cE}{\mathcal{E}}

\newcommand*{\cH}{\mathcal{H}}

\newcommand*{\cL}{\mathcal{L}}
\newcommand*{\cM}{\mathcal{M}}
\newcommand*{\cN}{\mathcal{N}}

\newcommand*{\cS}{\mathcal{S}}

\newcommand*{\ket}[1]{\left|#1\right\rangle}
\newcommand*{\bra}[1]{\left\langle #1\right|}
\newcommand*{\proj}[1]{\ket{#1}\bra{#1}}
\newcommand*{\Tr}{\textnormal{Tr}}

\newcommand*{\fr}[2]{\frac{#1}{#2}}

\newcommand*{\ee}{\mathsf e}
\newcommand*{\ii}{\mathsf i}

\newcommand{\nrm}[1]{\left\|#1\right\|}

\renewcommand{\d}[1]{\ensuremath{\operatorname{d}\!{#1}}}

\newcommand*{\assign}{\ensuremath{\kern.5ex\raisebox{.1ex}{\mbox{\rm:}}\kern -.3em =}}

\newcommand*{\rR}{\mathscr R}

\DeclareMathOperator*{\dint}{\int\!\!\!\!\int}

\hypersetup{
    pdftoolbar=true,        
    pdfmenubar=true,        
    pdffitwindow=false,     
    pdfstartview=FitH,    
}

\makeatletter
\begin{document}
\title{Phase-asymmetry resource interconversion via estimation}
\date{\today}
\author{Varun Narasimhachar}
\email{vnarasim@ucalgary.ca}
\affiliation{Institute for Quantum Science and Technology, University of Calgary, 2500 University Drive NW, Calgary, Alberta, Canada T2N 1N4}
\affiliation{Department of Physics and Astronomy, University of Calgary, 2500 University Drive NW, Calgary, Alberta, Canada T2N 1N4}
\author{Gilad Gour}
\affiliation{Institute for Quantum Science and Technology, University of Calgary, 2500 University Drive NW, Calgary, Alberta, Canada T2N 1N4}
\affiliation{Department of Mathematics and Statistics, University of Calgary, 2500 University Drive NW, Calgary, Alberta, Canada T2N 1N4}

\begin{abstract}
Practical implementations of quantum information-theoretic applications rely on phase coherences between systems. These coherences are disturbed by misalignment between phase references. Quantum states carrying phase-asymmetry act as resources to counteract this misalignment. In this paper, we construct protocols for interconverting these resources by first estimating the misalignment. Our method achieves sublinear, but otherwise arbitrarily high, asymptotic rate of conversion of pure resource states to mixed ones. No other method is known for asymptotic pure-to-mixed state conversion. Our method also has the advantage of achieving phase reference alignment in addition to resource conversion.
\end{abstract}

\maketitle

\section{Introduction}
All tasks and protocols in quantum information theory---quantum cryptography, quantum computation, quantum metrology, etc.---depend on reference frames for physical degrees of freedom, because the transformations and measurements involved in such tasks are defined relative to reference frames \cite{BRS07}. Reference frames associated with different subsystems involved in a task often become mutually misaligned. Such misalignment can be corrected for by exchanging quantum states carrying information about the relative alignment between the reference frames. This information is in the form of asymmetry of the state with respect to the degree of freedom associated with the reference frame.

The asymmetry contained in quantum states deteriorates and gets exhausted with use. For this reason, it is considered a resource. The resource theory of asymmetry characterizes and quantifies the resource of asymmetry, and formulates protocols to interconvert between various forms of asymmetry \cite{BRS03,JWBVP06,BRS07,GS08,MS13,MS11,Mar12}. It has also been established (see \cite{BRS07}, for example) that the imposition of a so-called ``superselection rule'' (SSR) with respect to a group $G$ of transformations (abbreviated $G$-SSR) is equivalent to lacking alignment with a reference frame with respect to $G$, and therefore leads to a resource theory of asymmetry with respect to $G$.

An example of such a situation is when two remote parties, Alice and Bob, share an optical communication channel but lack a common reference for the optical phase of the light. This can result due to phase noise \cite{Mos86}. In such a case, the only operations that they can perform on the mode of light are those that do not need information about the relative misalignment of their optical phase references. If Alice sends Bob many states that encode phase information, say coherent states of a certain amplitude, then Bob can use those states to estimate the misalignment, or directly as a quantum reference for the phase. In this sense, states containing asymmetry relative to the phase degree of freedom are resources.

The above example corresponds mathematically to an SSR relative to the group $U(1)$ (the circle group). Reference frames associated with $U(1)$ are required in all applications that depend on phase coherences in optical and atomic states, such as quantum cryptography \cite{Gis03} and quantum clock-synchronization \cite{Chu00}. We develop a protocol for converting an arbitrarily large number of copies of a ``source'' phase-asymmetry resource to copies of a ``target'' resource in the presence of an SSR. This task is known as asymptotic interconversion. Our protocol first uses the source copies as input in a phase estimation method, first developed in \cite{Chi11}, and used in \cite{SG12}. Thereafter, we prepare the target resource as though there were no SSR, using the newly-acquired alignment information. We call this method ``estimation-preparation''. Existing methods for asymptotic interconversion \cite{GS08,MS13,MS11,Mar12} try to interconvert resources directly without aligning reference frames. Our method has the advantage that it can convert pure resource states to mixed resource states, for which no direct transformation method is known. Furthermore, since our method uses phase estimation, it results in the acquisition of alignment information, in addition to preparing the target state. This information is classical and stays forever, facilitating future tasks.

Other than our main result on $U(1)$-SSR, we also apply the estimation-preparation method to the finite cyclic groups, $\bbZ_d$. This kind of SSR occurs, for example, when agents lack a reference frame with respect to chirality \cite{CDGMP05,STDS13}. In this case we are able to achieve arbitrarily high asymptotic rate of interconversion between resources. This is not in conflict with the linear interconversion rate result obtained in \cite{GS08} using direct transformation methods, since our figure of merit is fundamentally different from theirs.

This paper is organized as follows: In Section~\ref{estp}, we give a brief background of the relevant concepts, and lay out the estimation-preparation method. In Section~\ref{u1}, we state and prove the main result on $U(1)$-SSR. In Section~\ref{zd} we present the result on $\bbZ_d$-SSR. Finally, we give concluding remarks, including a comparison of the estimation strategy with direct transformation, in Section~\ref{concl}.

\section{Estimation-preparation method}\label{estp}
In quantum theory, superselection rules (SSR's) are certain constraints on the allowed dynamics, in addition to the fundamental selection rules that arise from the properties of atomic and nuclear systems \cite{BRS03,JWBVP06,BRS07}. An SSR is a rule forbidding any quantum evolution that violates certain symmetries. SSR's arise naturally in any context where a reference frame for a certain degree of freedom is missing. Another way to look at this connection between SSR's and reference frames is that access to appropriate reference frames can help in circumventing, partially or even completely, the imposition of SSR's.

Let $U:G\to\cL(\cH)$ be a unitary representation of a group $G$, acting on a Hilbert space $\cH$.\footnote{$\cL(\cH)$ denotes the set of linear operators on $\cH$.} Consider an SSR that allows only operations which commute with the action of $U(G)$. We refer to it as an SSR with respect to the group $G$, or $G$-SSR. Under a $G$-SSR, all states that are invariant under the action of $U(G)$ are considered free of cost:
$$\left\{\rho\in \cD(\cH): \forall g\in G,\;U(g)\rho U^\dagger(g)=\rho\right\}.$$
Here, by $\cD(\cH)$ we mean the set of all normalized density operators on $\cH$. All states outside the free set contain some amount of asymmetry, and are therefore resources. The allowed operations are all completely-positive (CP) maps $\cE:H_+(\cH)\to H_+(\cH)$ such that $\forall g\in G$,
$$U(g)\cE(\cdot)U^\dagger(g)=\cE\left(U(g)(\cdot) U^\dagger(g)\right),$$
that is, all CP maps whose action commutes with that of the unitary group $U(G)$. These are called $G$-covariant operations.

Viewing an SSR as a restriction on the allowed operations has been previously referred to as the ``constrained-dynamical'' perspective \cite{MS13}. We will take advantage of the fact that a $G$-SSR is mathematically equivalent to the lack of a reference frame for a degree of freedom associated with $G$. All our analysis will be based on a model of $G$-SSR where there exists a ``standard'' reference frame and an ``available'' reference frame. The ``correct'' orientation with respect to the $G$ degree of freedom is given by the standard frame, but the available frame is misaligned from this standard by a transformation $g_0\in G$ that is completely unknown to us. This has been called the ``information-theoretic'' perspective towards SSR's in \cite{MS13}, and has been formally presented, for example, in \cite{BRS07}.

In our information-theoretic perspective, we model the complete ignorance about $g_0$ by treating $g_0$ as a random variable distributed by the Haar measure on $G$ (the uniform distribution in the case of finite groups):
\begin{equation}\label{haar}
\Pr(g_0)=\left\{\begin{array}{ll}\d{g_0},&G\textnormal{ a compact Lie group,}\\\fr1{|G|},&G\textnormal{ a finite group.}\end{array}\right.
\end{equation}
Note that the measure $\d{g_0}$ is the group's invariant measure, and therefore independent of $g_0$; we use this notation merely as an index which will be convenient for summation.

In any quantum resource theory, the task of state interconversion encompasses the most general information-theoretic task one might have to perform: any task consists of taking as input some quantum and classical information, and producing as output some other quantum and classical information. In particular, we consider the task of asymptotic interconversion, which is to convert an arbitrarily large number of copies of a ``source'' resource state to copies of a desired ``target'' state:
$$\sigma^{\otimes N}\mapsto\tau^{\otimes M}.$$
The aim is to maximize the yield $M$ for a given $N$, as the latter approaches infinity. In fact, the requirement of exact conversion to copies of the output state is too strict and unnecessary. In asymptotic interconversion, we only require that there exist sequences $M(N)$ and $\tau^{(N)}$ such that
\begin{equation}
\lim_{N\to\infty}F\left(\tau^{(N)},\tau^{\otimes M(N)}\right)=1,
\label{asym}
\end{equation}
with the conversion $\sigma^{\otimes N}\mapsto\tau^{(N)}$ achievable for every $N\in\bbN$ using the allowed operations. Here $F(\cdot,\cdot)$ denotes the fidelity of a pair of density matrices.

In asymptotic interconversion, one is usually interested in finding such a sequence $M(N)$ whose terms get as large as possible asymptotically. However, we use a slightly, yet fundamentally, different notion of ``success'' at this task. As per our view of the $G$-SSR as the application of an unknown transformation $g_0\in G$, the target state is not a fixed state $\tau$, but rather a function of the random variable $g_0$:
$$\tau_{g_0}\assign U(g_0)\tau U^\dagger(g_0).$$
Our asymptotic conversion method, which we call ``estimation-preparation'', first produces an estimate $g$ for $g_0$, and then prepares copies of the state $\tau_{g}$. Instead of the condition (\ref{asym}), we require that
\begin{equation}
\lim_{N\to\infty}\dint\limits_{g_0,g\in G}F\left(\tau_{g}^{\otimes M(N)},\tau_{g_0}^{\otimes M(N)}\right)\Pr\left(g_0,g\right)=1,
\label{asest}
\end{equation}
where $\Pr\left(g_0,g\right)$ is the joint distribution of the true misalignment $g_0$ and estimate $g$. This distribution consists of the uniform distribution (\ref{haar}) of $g_0$ combined with the conditional distribution $\Pr(g|g_0)$, which is decided by the estimation protocol and the source state used as input. We use the estimation method of \cite{Chi11}, consisting of subjecting the source $\sigma^{\otimes N}$ to the measurement given by the POVM
\begin{equation}
E(B)=\int\limits_{g\in B}U(g)\proj\eta U^\dagger(g)\d g,
\label{covm}
\end{equation}
where $B\subset G$ is a measurable subset of $G$, $\d g$ is the Haar measure on $G$ (or just $1/|G|$ for a finite group), and $\ket\eta\in\cH$ is a special vector known as a class vector\footnote{We are taking $\cH$ large enough to accommodate spaces that carry tensor product copies of $U(G)$ as well.}. This is called the \emph{covariant measurement seeded by $\ket\eta$}. For the source state $\sigma^{\otimes N}$, the actual input to the measurement is $\sigma_{g_0}^{\otimes N}$ (due to the misalignment). The measurement results in the conditional distribution
$$\Pr(g|g_0)=\Tr\left(\sigma_{g_0}^{\otimes N}U(g)\proj\eta U^\dagger(g)\d{g}\right),$$
whence the joint distribution is
\begin{equation}
\Pr\left(g_0,g\right)=\bra\eta U(g^{-1}g_0)\sigma^{\otimes N}U^\dagger(g^{-1}g_0)\ket\eta\d g_0\d{g}.
\label{poste}
\end{equation}

In the following sections, we will apply this method to the groups $U(1)$ and $\bbZ_d$. The essential objective is as follows: for a given pair of source and target states $(\sigma,\tau)$, we want to find the best possible asymptotic scaling $M(N)$ such that (\ref{asest}) is satisfied for the distribution given by (\ref{poste}). We will restrict to pure source states, but the target is allowed to be mixed.

\section{$U(1)$-SSR}
\label{u1}
In this section we consider the case of an SSR with respect to the $U(1)$ group. Such an SSR arises, for example, when we lack a reference for the optical phase of a mode of light. A unitary representation of $U(1)$ can be decomposed into its one-dimensional irreducible representations (irreps), inducing a decomposition of the Hilbert space as $\cH\equiv\bigoplus_{n\in\bbZ}\cM_n\otimes\cN_n$, where each $\cM_n=\textnormal{Span}\left\{\ket n\right\}$ is $1$-dimensional and carries the irrep associated with $n$, while $\cN_n$ has dimension equalling the multiplicity of that irrep, and carries a trivial representation of $U(1)$. The action of the representation is given by $U(\theta)\left(\ket n\otimes\ket\alpha\right)=\ee^{\ii n\theta}\ket n\otimes\ket\alpha$, where $\ket\alpha$ is an arbitrary vector in $\cN_n$. The free states in the associated resource theory are all states that are $U(1)$-invariant, i.e. all states block-diagonal with respect to the index $n$ of the irrep.

For a pure state $\ket\psi=\sum_nc_n\ket n\otimes\ket{\alpha_n}$, we will call the set $\{n:c_n\ne0\}$ the \emph{number spectrum of $\ket\psi$}.

Our main result is the following.
\begin{theorem}\label{thu1}
Under a $U(1)$-SSR, using the estimation-preparation method, a pure resource state $\ket\phi$ can be converted to a pure or mixed resource state $\tau=\sum_1t_k\proj{\psi_k}$ at a sublinear, but otherwise arbitrarily high, asymptotic conversion rate, provided that $\ket\phi$ and the spectral components $\ket{\psi_k}$ of $\tau$ have gapless number spectra.
\end{theorem}
The remainder of this section consists of a proof of Theorem \ref{thu1}: the asymptotic rate at which pure and mixed target states can be obtained from \emph{pure} source states using the estimation-preparation method introduced in Section~\ref{estp}. We will prove the result first for pure target states.

\subsection{Pure target states}
\label{u1p}
We want to convert copies of some pure state to those of another pure state:
$$\ket\phi^{\otimes N}\mapsto\ket\psi^{\otimes M}.$$
Here we shall restrict consideration to pure states whose number spectra are bounded on at least one side. This restriction is naturally obeyed, for example, for the representation of $U(1)$ associated with the optical phase of a mode of light, but not for that associated with the gauge phase conjugate to electric charge. With this restriction, it is known from previous work \cite{GS08} that any pure state can be reversibly converted by $U(1)$-covariant operations to a ``standard'' state of the following form:
$$\ket\psi=\sum_{n=0}^\infty\sqrt{c_n}\ket n,$$
where $c_0>0$, $c_n\ge0$ and $\ket n\equiv\ket n\otimes\ket{\textnormal{Rep}_n}$ is a ``representative'' vector in the eigenspace corresponding to index $n$. Throughout this subsection, where we consider only pure states, we will restrict to the space $\cH\assign\bigoplus_{n=0}^\infty\textnormal{Span}\left(\ket n\right)$.

Let the input state be
$$\ket\phi=\sum_{n=0}^{m_1}\sqrt{p_n}\ket n$$
such that each $p_n\ne0$ in the sum (i.e. a gapless spectrum)\footnote{The case of pure states with number spectra without upper bound can be incorporated by later taking $m_1\to\infty$. This does not pose any convergence problems so long as the state is normalized. For clarity, we shall restrict to finite $m_1$, without loss of generality.}. In the limit of large $N$, the central limit theorem \cite{Gol10} guarantees that
$$\ket\phi^{\otimes N}=\sum_{n=0}^{Nm_1}\sqrt{\tilde P_n}\ket n\otimes\ket{\alpha_n}.$$
Here $\tilde{\mathbf P}$ is some distribution close to
\begin{equation}
P_n\assign\fr{\ee^{-\fr{\left(n-N\mu_\phi\right)^2}{2N\sigma_\phi^2}}}{\sqrt{2\pi N\sigma_\phi^2}},
\end{equation}
where $\mu_\phi$ is the mean and $\sigma_\phi^2$ the variance of the distribution $\mathbf p$. The error in the approximation can be bounded in the $\ell1$ distance between the distributions:
\begin{equation}
\nrm{\tilde{\mathbf P}-\mathbf P}_1=O\left(\fr1{\sqrt{N}}\right).
\label{err1}
\end{equation}
The multiplicity of each irrep would no longer be restricted to $1$ when the tensor product of copies of the original representation is decomposed. Therefore, instead of having a unique, state-independent $\ket n$ as before, we have here some $\ket n\otimes\ket{\alpha_n}$ where $\ket{\alpha_n}\in\cN_n$ depends on the single-copy state $\ket\phi$. However, we can again take this state to its standard form, ignoring the multiplicities.

In the case of $U(1)$, the seed $\ket\eta$ of the measurement $E$ shown in (\ref{covm}) can be chosen to be
\begin{equation}\label{seed}
\ket\eta\assign\sum_{n=0}^{Nm_1}\ket n.
\end{equation}
Note that this is equivalent to using the bipartite seed described in \cite{Chi11}, because all irreps of $U(1)$ are $1$-dimensional. The limits of the sum are chosen so that this POVM is complete on the support of $\proj\phi^{\otimes N}$. Let $\theta_0$ be the true misalignment. With the input state $\ket\phi^{\otimes N}$, we use (\ref{poste}) to obtain the posterior joint distribution of the misalignment and estimate:
\begin{align}
\Pr\left(\theta_0,\theta\right)=&\fr{\d{\theta_0}}{2\pi}\fr{\d{\theta}}{2\pi}\sum_{m,n}\ee^{\ii(m-n)\left(\theta-\theta_0\right)}\bra m\phi^{\otimes N}\ket n\nonumber\\
\approx&\fr{\d{\theta_0}}{2\pi}\fr{\d{\theta}}{2\pi}\sum_{m,n}\sqrt{P_mP_n}\ee^{\ii(m-n)\left(\theta-\theta_0\right)}\nonumber\\
\approx&\fr{\d{\theta_0}}{2\pi}\fr{\d\gamma}{\sqrt{2N\pi\sigma_\phi^2}}\left|\int\limits_{-\infty}^\infty\fr{\ee^{-\fr{\left(m-N\mu_\phi\right)^2}{4N\sigma_\phi^2}+\ii m\gamma}}{\sqrt{2\pi}}\d m\right|^2\nonumber\\
=&\fr{\d{\theta_0}}{2\pi}\d\gamma\sqrt{\fr{2N\sigma_\phi^2}{\pi}}\ee^{-2N\sigma_\phi^2\gamma^2},
\label{Posterior}
\end{align}
where $\gamma\assign\theta-\theta_0$, and we have used the arguments elaborated in \cite{SG12} to approximate the sum by an integral. The approximation in the second line incurs an $O\left(1/\sqrt{N}\right)$ error (see (\ref{err1})), which is also the order of the error associated with the third line \cite{SG12}. Therefore, the overall error is $O\left(1/\sqrt{N}\right)$.

Having used this estimation strategy, we wish to prepare an approximant to $M$ copies of some state $\ket\psi=\sum_n\sqrt{q_n}\ket n$, which we assume also has a gapless spectrum. For large $M$,
$$\ket\psi^{\otimes M}\approx\ket{\psi^{(M)}}\assign\sum_n\sqrt{Q_n}\ket n,$$
where
\begin{equation}
Q_n\assign\fr{\ee^{-\fr{\left(n-M\mu_\psi\right)^2}{2M\sigma_\psi^2}}}{\sqrt{2\pi M\sigma_\psi^2}},
\end{equation}
$\mu_\psi$ being the mean, and $\sigma_\psi^2$ the variance, of the distribution $\mathbf q$. Here the approximation error is $O\left(1/\sqrt{M}\right)$ in terms of the trace distance between distributions, as in (\ref{err1}).

Given outcome $\theta$ of our estimation POVM, we prepare the state
$$\ket{\psi^{(M)}_{\theta}}\assign\sum_n\ee^{\ii n\theta}\sqrt{Q_n}\ket n.$$
The fidelity of the prepared state with the true target state is
\begin{align}
F\left(\psi_{\theta_0}^{\otimes M},\psi_{\theta}^{\otimes M}\right)\approx&\left|\left\langle\psi^{(M)}_{\theta}\Big|\psi^{(M)}_{\theta_0}\right\rangle\right|^2\nonumber\\
=&\left|\sum_n\ee^{\ii n\left(\theta-\theta_0\right)}Q_n\right|^2\nonumber\\
\approx&\left|\int\limits_{-\infty}^\infty\fr{\d n}{\sqrt{2\pi M\sigma_\psi^2}}\ee^{\ii n\gamma}\ee^{-\fr{\left(n-M\mu_\psi\right)^2}{2M\sigma_\psi^2}}\right|^2\nonumber\\
=&\ee^{-M\sigma_\psi^2\gamma^2}.
\label{purf}
\end{align}
Both of the approximations above are to $O\left(1/\sqrt{M}\right)$, and so the overall error is $O\left(1/\sqrt{M}\right)$. We sum the above over the probability distribution (\ref{Posterior}) to obtain our figure of merit:
\begin{align}
f\left[\phi,N;\psi,M\right]\assign&\dint\limits_{\theta_0,\theta\in U(1)}F\left(\psi_{\theta_0}^{\otimes M},\psi_{\theta}^{\otimes M}\right)\Pr\left(\theta_0,\theta\right)\nonumber\\
\approx&\sqrt{\fr{2N\sigma_\phi^2}{\pi}}\dint\limits_{\theta_0,\gamma}\fr{\d{\theta_0}}{2\pi}\ee^{-M\sigma_\psi^2\gamma^2-2N\sigma_\phi^2\gamma^2}\d\gamma\nonumber\\
\approx&\fr{1}{\sqrt{1+\fr{M\sigma_\psi^2}{2N\sigma_\phi^2}}}
\label{purer}
\end{align}
for large $N$ and $M$. The errors accrued in the two approximations above are $O\left(1/\sqrt{N}+1/\sqrt{M}\right)$ and $O\left[\exp\left(-\pi^2\left(M\sigma_\psi^2+2N\sigma_\phi^2\right)^2\right)/\sqrt{M}\right]$, respectively. Therefore, the overall error is $O\left(1/\sqrt{M}\right)$. Note that $\sigma_\phi$ and $\sigma_\psi$ do not change with $N$: they are just properties of the single-copy states. Therefore, in order for $f$ to approach 1 asymptotically, $N$ must dominate $M$. The asymptotic expansion of $f$ is of the form
$$f\left[\phi,N;\psi,M\right]=1+O\left(\fr{M}{N}\right)+O\left(\fr1{\sqrt{M}}\right).$$

We see that any yield rate $M(N)$ that grows strictly sublinearly in $N$ allows $f$ to approach $1$ for large $N$. Therefore, we may choose any rate function $M(N)$ that grows strictly sublinearly in $N$.\qed

\subsection{Mixed target states}
\label{u1mt}
We now extend the proof in the previous section to the case of mixed target states. To this end, we use some results on typical sequences, which we provide first in Section~\ref{aep1}. Thereafter, in Section~\ref{mproof}, we prove Theorem \ref{thu1} for mixed target states.
\subsubsection{AEP applied to simplify tensor products}\label{aep1}
In the case where mixed states are involved, we must necessarily consider the multiplicity structure arising from a tensor product. Recall that the Hilbert space carrying the representation can be decomposed in the manner $\cH\equiv\bigoplus_n\cM_n\otimes\cN_n$, where $\cM_n=\textnormal{Span}\left\{\ket n\right\}$ is $1$-dimensional and carries the irrep associated with $n$, while $\cN_n$ has dimension equalling the multiplicity of that irrep, and carries a trivial representation of $U(1)$.

Consider a mixed resource state $\rho=\tau^{\otimes M}$, where
$$\tau=\sum_{k=1}^{R_\tau}t_k\proj{\psi_k}$$
is a spectral decomposition of $\tau$, with
$$\ket{\psi_k}=\sum_n\sqrt{p^{(k)}_n}\ket n\otimes\ket{\psi^{(k)}_n}.$$
We restrict to the case where each $\ket{\psi_k}$ has a gapless number spectrum.

For large $M$, $\rho$ can be approximated well by considering only the most typical pure states in its decomposition, i.e. those where $\ket{\psi_k}$ appears $\sim t_kM$ times in the tensor product, for each $k$. To make this notion precise, let us introduce a sequence of small real numbers $\epsilon\left(M\right)\ge0$ whose asymptotic limit is zero, but whose convergence to the limit is ``slow enough'':
$$\epsilon\left(M\right)=o\left(M^0\right),$$
$$\sqrt\fr{\log M}{M}=o\left(\epsilon\left(M\right)\right).$$
For any $\epsilon\ge0$, define the set of $R_\tau$-dimensional probability distributions
$$\rR_\epsilon\assign\left\{\textnormal{Probability distributions }\tilde{\mathbf t}\in\bbR^{R_\tau}:\nrm{\tilde{\mathbf t}-\mathbf t}_1\le\epsilon\right\},$$
where $\nrm\cdot_1$ is the $\ell1$ distance between probability distributions. The asymptotic equipartition property (AEP) of strongly-typical sequences \cite{Sha01} allows us to restrict the expansion of $\tau^{\otimes M}=\left(\sum_{k=1}^{R_\tau}t_k\proj{\psi_k}\right)^{\otimes M}$ to only those terms where each $\ket{\psi_k}$ appears $\tilde t_kM$ times in the tensor product for $\tilde{\mathbf t}\in\rR_{\epsilon\left(M\right)}$, incurring an overall error in trace norm of $O\left(\epsilon\left(M\right)\right)$:
\begin{equation}\label{typ}
\rho=\delta_\rho\rho^{\textnormal{res}}+\sum_{j=1}^{R_\rho}r_j\proj{\beta^{(j)}}
\end{equation}
with $0\le\delta_\rho=O\left(\epsilon\left(M\right)\right)$, $r_j\ge0$, and $\rho^{\textnormal{res}},\beta^{(j)}$ normalized density operators.

Here, for each $j\in\{1,\dots,R_\rho\}$,
$$\ket{\beta^{(j)}}=U_{\pi_j}\bigotimes_{k=1}^{R_\tau}\ket{\psi_k}^{\otimes t^{(j)}_kM},$$
where $\mathbf t^{(j)}\in\rR_{\epsilon\left(M\right)}$, and $U_{\pi_j}$ is a unitary that permutes the $M$ subsystems by some permutation $\pi_j\in S_M$. Now, for each $k\in\{1,\dots,R_\tau\}$,
$$\ket{\psi_k}^{\otimes t^{(j)}_kM}=\sum_n\sqrt{\tilde P^{(j,k)}_n}\ket n\otimes\ket{\psi^{(j,k)}_n},$$
where $\nrm{\tilde{\mathbf P}^{(j,k)}-\mathbf P^{(j,k)}}_1=O(1/\sqrt M)$ with
\begin{align}
P^{(j,k)}_n&=\fr{\ee^{-\fr{\left(n-t^{(j)}_kM\mu_{\psi_k}\right)^2}{2t^{(j)}_kM\sigma_{\psi_k}^2}}}{\sqrt{2\pi t^{(j)}_kM\sigma_{\psi_k}^2}},
\end{align}
We can now find the number distribution in $\ket{\beta^{(j)}}$ using the fact that the sum of independent normally-distributed variables is also normally distributed. Defining
$$P^{(j)}_n\assign\fr{\ee^{-\fr{\left(n-M\mu^{(j)}_\tau\right)^2}{2M\left(\sigma^{(j)}_\tau\right)^2}}}{\sqrt{2\pi M\left(\sigma^{(j)}_\tau\right)^2}}$$
with $\mu^{(j)}_\tau\assign\sum_{k=1}^{R_\tau}t^{(j)}_k\mu_{\psi_k}$ and $\sigma^{(j)}_\tau\assign\sqrt{\sum_{k=1}^{R_\tau}t^{(j)}_k\sigma_{\psi_k}^2}$, we have
$$\ket{\beta^{(j)}}=\sum_n\sqrt{\tilde P^{(j)}_n}\ket n\otimes\ket{\beta^{(j)}_n}$$
for some $\ket{\beta^{(j)}_n}\in\cN_n$ and $\nrm{\tilde{\mathbf P}^{(j)}-\mathbf P^{(j)}}_1=O(1/\sqrt M)$.

\subsubsection{Proof for mixed target states}\label{mproof}
Now let the state $\rho=\tau^{\otimes M}$ of Section~\ref{aep1} be the target state of the estimation-preparation protocol. For a true misalignment $\theta_0\in U(1)$ and an estimate $\theta$, using (\ref{typ}), the fidelity between the true target and prepared target states would be
\begin{align}
F\left(\tau_{\theta_0}^{\otimes M},\tau_{\theta}^{\otimes M}\right)\ge&\delta_\rho F\left(\rho_{\theta_0}^{\textnormal{res}},\rho_{\theta}^{\textnormal{res}}\right)+\sum_{j=1}^{R_\rho}r_jF\left(\beta_{\theta_0}^{(j)},\beta_{\theta}^{(j)}\right)\nonumber\\
\ge&\left(1+o(M^0)\right)\min_jF\left(\beta_{\theta_0}^{(j)},\beta_{\theta}^{(j)}\right),
\end{align}
where we have used the joint concavity of the fidelity function in its arguments, and the fact that $\delta_\rho=O\left(\epsilon(M)\right)=o(M^0)$.

For each $j$ the state $\beta^{(j)}$ is pure, and therefore $F\left(\beta_{\theta_0}^{(j)},\beta_{\theta}^{(j)}\right)$ can be calculated by the same method as in Section~\ref{u1p} (cf. (\ref{purf})), yielding
$$F\left(\beta_{\theta_0}^{(j)},\beta_{\theta}^{(j)}\right)=\ee^{-M\left(\sigma^{(j)}_\tau\theta\right)^2}+O\left(\fr1{\sqrt M}\right),$$
where $\theta=\theta-\theta_0$ and $\sigma^{(j)}_\tau\assign\sqrt{\sum_{k=1}^{R_\tau}t^{(j)}_k\sigma_{\psi_k}^2}$. Let the minimum be attained by $j=j_0$. Then, we have
$$F\left(\tau_{\theta_0}^{\otimes M},\tau_{\theta}^{\otimes M}\right)\ge\ee^{-M\left(\sigma^{(j_0)}_\tau\theta\right)^2}+O\left(\fr1{\sqrt M}\right).$$

Since the source is a pure state $\phi^{\otimes N}$, we use the same method as in (\ref{purer}), together with the above bound on the fidelity, to obtain the bound
\begin{align}
f[\phi,&N;\tau,M]\nonumber\\
&\ge\fr{1}{\sqrt{1+\fr{M\left(\sigma^{(j_0)}_\tau\right)^2}{2N\sigma_\phi^2}}}+O\left(\fr1{\sqrt M}\right)+O\left(\fr1{\sqrt{N}}\right)\nonumber\\
&=1+O\left(\fr{M\left(\sigma^{(j_0)}_\tau\right)^2}{N}\right)+O\left(\fr1{\sqrt M}\right)+O\left(\fr1{\sqrt{N}}\right).
\end{align}
But since $\mathbf t^{(j_0)}\in\rR_{\epsilon\left(M\right)}$, $\sigma^{(j_0)}_\tau=\sigma_\tau+O\left(\epsilon\left(N\right)\right)$, where $\sigma_\tau\assign\sqrt{\sum_{i=1}^{R_\tau}t_k\sigma_{\psi_k}^2}$. This leaves us with
\begin{align}
f\left[\phi,N;\tau,M\right]\ge1&+O\left(\fr{M}{N}\right)+O\left(\fr{M\epsilon\left(M\right)}{N}\right)\nonumber\\
&+O\left(\fr1{\sqrt{M}}\right)+O\left(\fr1{\sqrt{N}}\right).
\end{align}
Therefore, we again recover the result that perfect asymptotic conversion is achievable at any rate $M(N)$ that grows sublinearly in $N$.\qed

\section{$\bbZ_d$-SSR}
\label{zd}
Here we consider an SSR with respect to a unitary representation $U$ of $\bbZ_d$. Such a representation can be generated by $U_1$ such that $\forall j\in\bbZ_d$, $U(j)=(U_1)^j$. It acts on a $d$-dimensional Hilbert space $\cH_d\assign\textnormal{Span}\left\{\ket j\right\}_{j\in\bbZ_d}$, where $\forall j,k\in\bbZ_d$,
$$\bra jk\rangle=\delta_{jk}$$
and
$$U_1\ket j=\omega^j\ket j,$$
with $\omega\assign\ee^{\ii\fr{2\pi}d}$.

The case of $\bbZ_d$ is very different from that of $U(1)$: in this case, there exist quantum states whose single copy can completely obviate the SSR. For example, let the unknown misalignment be $m\in\bbZ_d$, and suppose that we are given just one copy of the state $\ket{\eta_0}\assign\fr1{\sqrt d}\sum_{j\in\bbZ_d}\ket j$ from the source. Due to the interjection of the unknown $m$, the state in our reference frame is
$$U(m)\ket{\eta_0}=:\ket{\eta_{m}}=\fr1{\sqrt d}\sum_{j\in\bbZ_d}\omega^{mj}\ket j.$$
If we perform a von Neumann measurement in the basis $\left\{\ket{\eta_j}\right\}_{j\in\bbZ_d}$ (which is always an orthonormal basis spanning $\cH_d$)\footnote{This measurement is in fact the coherent measurement seeded by $\ket{\eta_0}$ (see (\ref{covm})).}, then we get the correct value $m$ with certainty. Thereafter, this perfect estimate may be used to prepare any state.

Since just one copy of this state suffices, using one such copy, we can prepare any state at an infinite rate (and even restore the copy we used in the measurement). However, if no copy of such an ``ultimate resource'' is available, but we have many copies of a different \emph{pure} state, then what rate can be achieved?

The groups $\bbZ_d$ have the property that for $k$ copies $\left\{U_{(i)}\right\}_{i=1}^k$ of the representation $U$,
$$\bigotimes_{i=1}^kU_{(i)}\cong U\otimes T_{d^{k-1}},$$
where $T_{d^{k-1}}$ is an $d^{k-1}$-dimensional trivial representation of $\bbZ_d$. The Hilbert space carrying the direct product representation likewise becomes the direct product of one copy of $\cH_d$ and a multiplicity space carrying the trivial representation. Any transformation within the multiplicity space is free. Therefore, since we consider only pure states, we may replace any pure state on the product space by a ``canonical representative'' state on $\cH_d$.

Suppose now that we are given $N$ copies of a state $\ket\psi\assign\sum_{j\in\bbZ_d}\sqrt{p_j}\ket j$. These copies can be replaced by the representative
$$\ket{\psi_{N}}=\sum_{j\in\bbZ_d}\sqrt{c_j}\ket j,$$
where
$$c_j\assign\sum_{\mathbf t\in\cS(N,j)}\left(\begin{array}{c}N\\\mathbf t\end{array}\right)\prod_{k\in\bbZ_d}p_k^{t_k},$$
with $\cS(N,j)$ the collection of all $d$-tuples $\mathbf t$ of nonnegative integers such that $\sum_{k\in\bbZ_d}t_k=N$ and $\sum_{k\in\bbZ_d}kt_k\cong j\mod d$. From the analysis shown in \cite{SG12}, it follows that for any $\mathbf p$ with at least $2$ nonzero components\footnote{That is, any $\mathbf p$ such that $\nexists j:p_j=1$.},
\begin{equation}\label{asflat}
c_j=\fr1d+O\left(\epsilon^N\right),
\end{equation}
with $0\le\epsilon<1$.

Now, if we measure this state using the measurement mentioned before, namely the von Neumann projective measurement in the basis $\left\{\ket{\eta_j}\right\}_{j\in\bbZ_d}$, then the probability of getting a particular wrong outcome $m_1\ne m$ is
$$\Pr\left(m_1|m\right)=\left|\bra{\eta_{m_1}}U(m)\ket{\psi_N}\right|^2=\left|\sum_j\omega^{(m_1-m)j}\sqrt{c_j}\right|^2.$$
Using (\ref{asflat}),
$$\sqrt{c_j}=\fr1{\sqrt d}+O\left(\epsilon^N\right),$$
whence
\begin{align}
\Pr\left(m_1|m\right)=&\left|\sum_j\omega^{(m_1-m)j}\left(\fr1{\sqrt d}+O\left(\epsilon^N\right)\right)\right|^2\nonumber\\
=&\fr1d\left|\sum_j\omega^{(m_1-m)j}+O\left(\epsilon^N\right)\right|^2\nonumber\\
=&O\left(\epsilon^{2N}\right),
\end{align}
where we have used the fact\footnote{Note that this is true for composite $d$ as well.} that $\sum_j\omega^{(m_1-m)j}=0$ for any $m_1\ncong m\mod d$.

Therefore, the probability of getting the correct outcome is bounded below as
$$\Pr\left(m|m\right)=1-\sum_{m_1\in\bbZ_d\setminus\{m\}}\Pr\left(m_1|m\right)=1-O\left(\epsilon^{2N}\right).$$

Since the fidelity of preparation of any target state is perfect whenever the outcome is correct, this gives us trivially that the average fidelity (over all $m$) is also bounded from below by the same quantity. Because this bound depends only on the number of input copies and not on the output state or the number of output copies, in the asymptotic limit we may choose any preparation rate $M(N)$, without bound, for pure or mixed target states.

Note that this result on $\bbZ_d$-SSR is not in conflict with our result on $U(1)$-SSR (Section~\ref{u1}), where our method achieves a sublinear rate. The two cases are fundamentally different: our method in the $\bbZ_d$ case relies on being able to make the number of copies $N$ arbitrarily large, not just absolutely, but also in particular relative to $d$. On the other hand, in the $U(1)$ case we were able to appeal to the law of large numbers to approximate certain probability distributions by Gaussian distributions, and to extend the domain of the sample space to all real numbers, only by assuming that this domain is, as such, much wider than the scale of $N$. Only after this step do we allow $N$ itself to grow to infinity.

\section{Discussion and conclusion}
\label{concl}
In this paper, we formulate an estimation-preparation strategy for asymptotic state interconversion in the presence of a symmetry superselection rule (SSR). Estimation-preparation consists of first estimating the unknown misalignment of the reference frame using the methods described in \cite{Chi11}, and thereafter using the obtained alignment information to prepare the target state.

In the case of an SSR associated with a phase ($U(1)$-SSR), our estimation strategy can achieve a sublinear, but otherwise arbitrarily high, rate of conversion asymptotically, as long as the source state is pure and has a gapless number spectrum. In \cite{GS08}, a direct transformation strategy has been constructed to convert pure states to pure states. This direct strategy achieves a linear conversion rate, compared to the sublinear rate of our method. However, our method performs equally well in the case where the target state is mixed, whereas no direct strategy is known in this case.

Furthermore, since our method includes a step where the reference frame misalignment is estimated, the result of the protocol is not only the preparation of the desired target state, but also an alignment of the reference frame. This alignment is in the form of classical information, and therefore, can be reused in future tasks. In direct transformation strategies, on the other hand, the misalignment remains even after state conversion.

For an SSR associated with one of the finite cyclic groups, $\bbZ_d$, our strategy can achieve arbitrarily high asymptotic conversion rate when the source state is pure. This does not conflict with the result of \cite{GS08} that the optimal rate is linear: their figure of merit is the fidelity of the prepared state and a fixed target state, under constrained dynamics, whereas ours is the average of the fidelity over all possible target states (corresponding to all possible misalignments).

There is scope for future research in the problem of extending the estimation method to allow efficient state preparation from mixed sources. It also remains to determine whether the measurement we have used is optimal for estimation-preparation (although it is known to be optimal for estimation alone, under some conditions \cite{Chi11}). Another future direction is the application of our method to SSR's associated with more complex groups such as $SU(2)$ (spatial rotation-symmetry SSR) and $S_n$ (particle-exchange SSR). Finally, there remains the problem of resource conversion in the non-asymptotic (single-shot) regime, where finite copies of resources are considered. This case would not admit such convenient mathematical simplifications as the central limit theorem and asymptotic equipartition property, and presents a challenging problem for future work.

The authors thank Iman Marvian, Giulio Chiribella and Robert W. Spekkens for helpful discussions.

\bibliographystyle{ieeetr}
\bibliography{./Paper_estimation_formation}
\end{document}